\documentstyle[prb,aps,preprint]{revtex}
\begin{document}
\title{Electron-Electron Interaction in Linear Arrays of Small Tunnel
Junctions}
\author{K. K. Likharev and K. A. Matsuoka}
\address{Department of Physics, State University of New York, Stony
Brook, NY 11794-3800}

\maketitle

\begin{abstract}
We have calculated the spatial distribution of the electrostatic
potential created by an unbalanced charge $q$ in one of the conducting
electrodes of a long, uniform, linear array of small tunnel junctions.
The distribution describes, in particular, the shape of a topological
single-electron soliton in such an array.  An analytical solution
obtained for a circular cross section model is compared with results
of geometrical modeling of a more realistic structure with square
cross section.  These solutions are very close to one another, and can
be reasonably approximated by a simple phenomenological expression.
In contrast to the previously accepted exponential approximation, the
new result describes the crossover between the linear change of the
potential near the center of the soliton to the unscreened Coulomb
potential far from the center, with an unexpected ``hump'' near the
crossover point.
\end{abstract}


\narrowtext
\newpage
Recent theoretical and experimental studies have resulted in
considerable progress in understanding correlated single-electron
transfer in ultra-small tunnel junctions (for reviews, see
Refs. \onlinecite{ibm,mes,sct}).  These phenomena may be used as a
background for a new generation of analog and digital devices.  The
most common component of single-electronic devices is a
one-dimensional array of small tunnel junctions
(Fig. \ref{circfig}(a)).  Thus it is very important to gain a
quantitative understanding of the Coulomb interaction potential,
$U(r)$, between single electrons in such an array.

To our knowledge, all previous works on this topic (see, e.g.,
Ref. \onlinecite{dels} and references therein) have used a simple
model in which the complete matrix $[C]$ of mutual capacitances
between conducting ``islands'' of the array is truncated to
tridiagonal form.  In this form of the matrix, the only non-zero
elements are (a) the diagonal elements $C_{i,i} = C_o$, representing
the stray capacitances of the islands, and (b) the nearest-neighbor
elements $C_{i,i\pm 1} = C$, dominated by tunnel junction
capacitances.  Electron-electron interaction in the tridiagonal model
is described by a simple exponential law \cite{ibm,magn}:

\begin{equation}
U_t(r)=U_t(0)\exp (-m/m_o), 			\label{tradit}
\end{equation}

\noindent where $m = r/a$ is the distance between the two
electrons, in units of the array period $a$ (i.e., in number of
islands).  The parameters $U_t(0)$ and $m_o$ depend on the $C/C_o$
ratio, and in the most important limit of $C_o \ll C$:

\begin{eqnarray}
U_t(0)&=&\frac{e^2}{2\sqrt{CC_o}}, 		\label{fo} \\
m_o&=&\sqrt{\frac{C}{C_o}}.			\label{mo}
\end{eqnarray}

The tridiagonal model is strictly correct only if the array is placed
parallel to, and very near, a conducting ground plane.  However, we
are not aware of any experiments which actually have used such a
configuration.  The presence of a ground plane would increase the
stray capacitance $C_o$, thus cutting off the single-electron soliton
radius $m_o$ and suppressing the electron-electron interaction at
large distances.  Thus, it was our goal to describe long linear arrays
of realistic geometry without a ground plane.

If the length scale of the electron-electron interaction within the
array is much larger than the array period (i.e., if $m_o\gg 1$), the
electrostatics of the array should not depend strongly on the details
of the geometry of its islands.  Thus we can model the experimental
array geometry, such as the one shown in Fig. \ref{circfig}(a), with
an array of cylindrical islands of arbitrary cross section, such as
the ones shown in Fig. \ref{circfig}(b) and Fig. \ref{squarefig}(a),
provided that we keep the same junction capacitance $C=\epsilon S/4\pi
d$ and array period $a$.

To bring the interaction problem to an analytically calculable form,
we can investigate the continuum limit in which the discrete periodic
structure is replaced by a continuous dielectric medium (e.g., as
shown in Fig. \ref{circfig}(c)).  This approximation is valid when the
characteristic length of electron-electron interaction within the
array is much larger than the array period, i.e. $m_o\gg 1$.  The
effective dielectric constant $\epsilon_{ef}$ of the medium can be
found from the requirement that the relation between the average
electric field $\langle E\rangle=(1/a)\int ^a_0 E\,dz$ and average
displacement $\langle D\rangle$ is the same in the dielectric model as
in the array model:

\begin{displaymath}
\epsilon_{ef}=\frac{a}{d}\epsilon=\frac{4\pi aC}{S}\,.
\end{displaymath}

For the particular case of the dielectric model with circular cross
section of radius $R = \sqrt{S/\pi}$ (Fig. \ref{circfig}(c)), we can
find the electron-electron interaction energy $U_d(r)$ from the
electrostatic potential $\phi(\rho,z)$ induced in the dielectric
cylinder by a charge $e$ located at $z = 0$:

\begin{equation}
U_d(r) = e\phi(0,z=r).				\label{onemeqn}
\end{equation}

For the relatively large distances we are interested in, $m\sim m_o\gg 1$
\mbox{(i.e. $ma\gg R$)} the shape of the initial charge is not important,
and it is natural to spread it uniformly over a thin disk ($z=0,
\rho\le R$).  The resulting boundary electrostatics problem can be
readily solved by the standard Fourier integral expansion:

\begin{equation}
\phi(0,z)=\cases{\phi(0,0) + \frac{e}{R^2}\left[
		\int^\infty_0 I_o(k\rho)\,A(k)\cos(kz)\,dk
			- \frac{2}{\epsilon_{ef}}|z|\right]\,, &
			\mbox{\quad $\rho \le R$\,,}\cr
		\int^\infty_0 K_o(k\rho)\,B(k)\cos(kz)\,dk\,, &
			\mbox{\quad $\rho \ge R$\,,}\cr }
\label{phioz}
\end{equation}

\noindent where the last term in the upper line takes care of the boundary
condition (Gauss's law) on the charged disk at $z=0$.  On the other
hand, the boundary conditions at $\rho=R$\,
($\partial\phi/\partial z\vert_+= \partial\phi/\partial z\vert_-,\,\,
\partial\phi/\partial\rho\vert_+ = \epsilon_{ef}\,
\partial\phi/\partial\rho\vert_-$) give

\begin{equation}
A(k)=\frac{4}{\pi \epsilon_{ef}k^2}
\left[I_o(kR)+\epsilon_{ef}\frac{I_1(kR)K_o(kR)}{K_1(kR)}\right]^{-1}.
\label{wavenum}
\end{equation}

\noindent [For numerical calculation of the integrals in Eqs. (\ref{phioz}),
(\ref{wavenum}), it is convenient to use the expansion $|z|\equiv
(2/\pi) \int ^\infty_0 \cos(kz)\,k^{-2}\,dk$, in order to cancel the
divergence of $A(k)$ at $k\to 0$.]

Numerical integration yields the functions $U_d(r)$ as shown in Fig.
\ref{plotfig}.  At large distances, these functions approach the free
space Coulomb interaction:

\begin{equation}
U_d(r) \to \frac{e^2}{r}\,,
\mbox{\qquad$r\to\infty$\,,}		\label{largedist}
\end{equation}

\noindent while at small distances, these functions are linear:

\begin{eqnarray}
U_d(r)&\to&U_d(0) - eE(0)r,\mbox{\qquad$r\to0$\,,} \label{bound} \\
\nonumber E(0)&=&\frac{2e}{\epsilon_{ef}R^2}\,.
\end{eqnarray}

\noindent The crossover between these two limits takes place at $r
\simeq r_o \equiv R\sqrt{\epsilon_{ef}}$.  Quite unexpectedly, at $r
\simeq r_o$ the potential $U_d(r)$ is {\em higher} than the
asymptotic value (\ref{largedist}), approaching it at $r\to\infty$
from {\em above}.

If $r_o \gg a$, we can apply this result to the discrete array of a
circular cross section.  In terms of the number of islands, the
crossover point is

\begin{equation}
m_o=\frac{r_o}{a}=\frac{R}{a}\sqrt{\epsilon_{ef}}=\sqrt{\frac{4C}{a}}\,.
\label{newmo}
\end{equation}

\noindent If we define the stray capacitance per unit period of the array as

\begin{equation}
C_o \equiv \frac{a}{4},			 \label{coexp}
\end{equation}

\noindent we can once again formally express $m_o$ as in the
tridiagonal model, $m_o = \sqrt{\frac{C}{C_o}}$.

The dielectric-model potential function $U_d$ can be approximated reasonably
well by a simple phenomenological expression:

\begin{equation}
U_a(m) = \frac{e^2}{a} \left\{
\frac{\alpha}{m_o} \exp(\frac{-\kappa m}{m_o})+
\frac{1}{m}\left[1-\exp(\frac{-\kappa m}{m_o})\right] \right\}\,.
\label{simpexp}
\end{equation}

\noindent with two dimensionless fitting parameters, $\alpha$ and
$\kappa$.  The boundary condition (\ref{bound}), together with the
definition for $m_o$ given in Eq. (\ref{newmo}), yields the condition
$\alpha = 2/\kappa -\kappa /2$, leaving only one free parameter,
$\kappa$.  Figure\ \ref{fig4} shows the dependence of $\kappa$ on
$m_o$, when $\kappa$ is adjusted to provide the best fitting (by
sight) of $U_a(m)$ to $U_d(m)$ (see Fig. \ref{plotfig}).  One can see
that $\kappa$ is close to unity, and depends on $m_o$ only
logarithmically, so that

\begin{displaymath}
U_a(0) = \frac{e^2(\alpha+\kappa)}{m_o a} \simeq
\frac{2e^2}{a}\sqrt{\frac{C_o}{C}} = \frac{e^2}{2\sqrt{C C_o}}.
\end{displaymath}

Thus, $U_a(0)$ may be approximated by $U_t(0)$ (Eq. (\ref{fo})) if
$m_o$ is within the practical interval $\sim 3-10$ (see below).

In order to check the validity of our results for structures with a
different cross section, we have calculated the interaction energy
$U_s(m)$ for a chain of square cross section islands (Fig.\
\ref{squarefig}(a)) for several values of the $a/d$ ratio using the
geometric capacitance modeling program FASTCAP \cite{fast}.  This
program takes, as input, a collection of ``panels,'' finite elements
representing the surfaces of a group of conductors.  By calculating
the amount of charge induced on each panel when one conductor is held
at fixed potential, the rest at zero, FASTCAP calculates, one by one,
each row of the capacitance matrix for the group of conductors.

For FASTCAP to calculate the capacitances accurately, the
discretization of the surface must be fine enough to represent the
charge distribution.  The main criterion, when panels (and computer
memory) are limited, is to let the panelling reflect the increased
charge density near the corner of a conductor.  Thus, when modeling
conductors with square cross section, each face was divided into 9
panels (Fig. \ref{squarefig}(b)), with smaller panels along the edges
where charge density is higher.  To test the accuracy of this simple
panelling, we modeled a chain of 10 islands in two ways: one with 9
panels per face, the other with 100 per face.  Differences in
electron-electron interaction energies between the two models were
less than 1\%, leading us to believe that the 9-panel model is
sufficiently accurate for our purposes in this work.

Island chains modeled with FASTCAP were limited in length to around
$120$ islands, due to computer memory limitations.  To avoid edge
effects, the energy of interaction between two electrons was
calculated with both electrons separated from the edges of the array
by at least $\sim 2 m_o$ islands.  Energies calculated this way did
not depend significantly on the total number of islands in the chain.
The results of the FASTCAP calculations are shown as points in Fig.
\ref{plotfig}.  One can see that they are very close to those obtained by
the dielectric model, and hence to the analytical approximation of Eq.
(\ref{simpexp}), despite a substantial difference in the geometry of
the two cross sections.

To summarize, we have shown that electron-electron interactions in
long one-dimensional arrays of small tunnel junctions are much better
described by Eq.(\ref{simpexp}) (with $\kappa \simeq 1$,\, $\alpha
\simeq 2$, and $U_a(0) \simeq e^2/2\sqrt{C C_o}$) than by the
traditional expression (\ref{tradit}).  The effective radius
$m_o=\sqrt{C/C_o}$ of the interaction is determined by the ratio of
the tunnel junction capacitance $C$ to the effective stray capacitance
$C_o$ (\ref{coexp}).  For an array formed on a substrate with
dielectric constant $\epsilon_{s}$, Eq. (\ref{coexp}) should be
modified to

\begin{displaymath}
C_o \simeq \frac{(\epsilon_{s} + 1)a}{8} .
\end{displaymath}

For a typical present-day single-electronic array with $a \simeq 0.1
\mu m$, formed on a SiO$_2$ substrate (see, e.g., Ref.
\onlinecite{haus}), the formula above gives $C_o/a \simeq 10^{-16}F/\mu m$
and $C_o \simeq 10^{-17}F$.  Thus, for the typical junction
capacitance $C \simeq 10^{-16}F$, the soliton radius $m_o$ is close to
3.  This estimate shows that for single-electronic devices where a
considerable soliton radius is important (for example, for suppression
of the macroscopic quantum tunnelling \cite{mes,sct}), vertical
structures with stacked junctions may be more advantageous.  Vertical
stacking can provide a very small array period $a$, and hence the
small stray capacitance $C_o$ needed for $m_o \gg 1$.

The authors are grateful to J. White and K. Nabors for providing the
program FASTCAP, and to D. Averin, A. Korotkov and J. Lukens for
numerous discussions.  This work was supported by AFOSR Grant
No. 91-0445.

\begin{figure}
\caption{One-dimensional array of tunnel junctions: (a) a typical
experimental geometry, with tunnel junctions of area $S$ and permittivity
$\epsilon$; (b) cylindric island model; and (c) dielectric
approximation to cylindric islands.}
\label{circfig}
\end{figure}

\begin{figure}
\caption{(a) Square cross section island array model. (b) Boundary
elements (``panels'') for square cross section island model used for
in FASTCAP capacitance calculations.}
\label{squarefig}
\end{figure}

\begin{figure}
\caption{Functions $U(m)$ describing the shape of the single electron
soliton.  Solid lines: $U_d(m)$, dielectric approximation to circular
cross section islands, with $\protect\epsilon_{ef} = 30,100,300$
(bottom to top).  To compare numerical and analytical results, we have
set $\pi R^2 = a^2$, so that $m_o=
\protect\sqrt{\protect\epsilon_{ef}/\pi}$.  Squares: $U_s(m)$, square cross
section array for the case $a=b$ (see Fig. \protect\ref{squarefig}).
Dashed lines: $U_a(m)$, as given by Eq. (\protect\ref{simpexp}).
Dotted line: free space potential, $U(m)=e^2/ma$.  Note that a vertical
offset is used in plotting the curves.}
\label{plotfig}
\end{figure}

\begin{figure}
\caption{The best-fit parameter $\kappa$\,, the sum $\kappa + \alpha$,
and $U(0)$, as functions of $m_o$\,.  Squares: $U_s(0)$.  Open
circles: $U_d(0)$.  Diamonds: $U_a(0)$.  Error bars reflect the range
over which a ``good'' fit, by sight, is achieved.}
\label{fig4}
\end{figure}


\newpage

\begin{references}

\bibitem{ibm}K. K. Likharev, IBM J. Res. Devel. {\bf 32}, 144 (1988).

\bibitem{mes}D. V. Averin and K. K. Likharev, in {\it Mesoscopic Phenomena in
Solids}, edited by B. Altshuler, P. A. Lee, and R. A. Webb (Elsevier,
Amsterdam, 1991), p. 173.

\bibitem{sct}{\it Single Charge Tunneling}, edited by H. Grabert and M. H.
Devoret (Plenum, New York, 1992).

\bibitem{dels}P. Delsing, in Ref. \onlinecite{sct}, p. 249.

\bibitem{magn}K. K. Likharev, N. S. Bakhvalov, G.. S. Kazacha, and S. I.
Serdyukova, IEEE Trans. Magn. {\bf 25}, 1436 (1989).

\bibitem{fast}K. Nabors, S. Kim, and J. White, IEEE Trans. on
Microwave Theory and Techniques, {\bf 40}, 1496 (1992).

\bibitem{haus} P. D. Dresselhaus, L. Ji, S. Han, J. E. Lukens, and K.
K. Likharev, Phys. Rev. Lett. {\bf 72}, 3226 (1994).
\end{references}
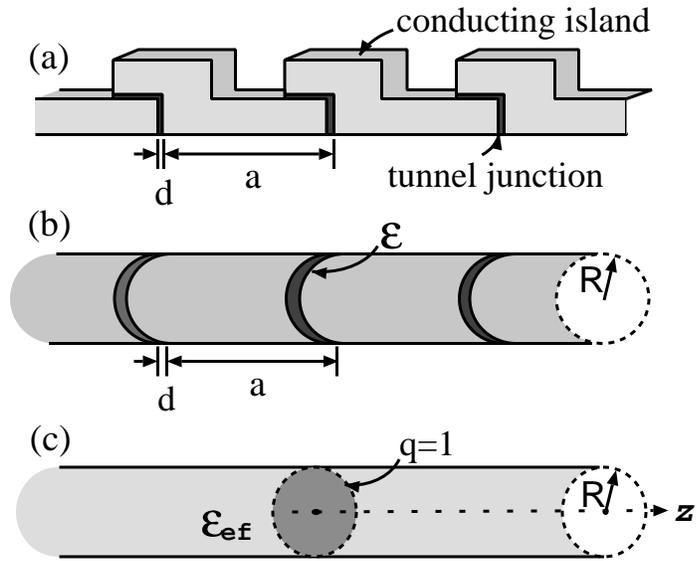
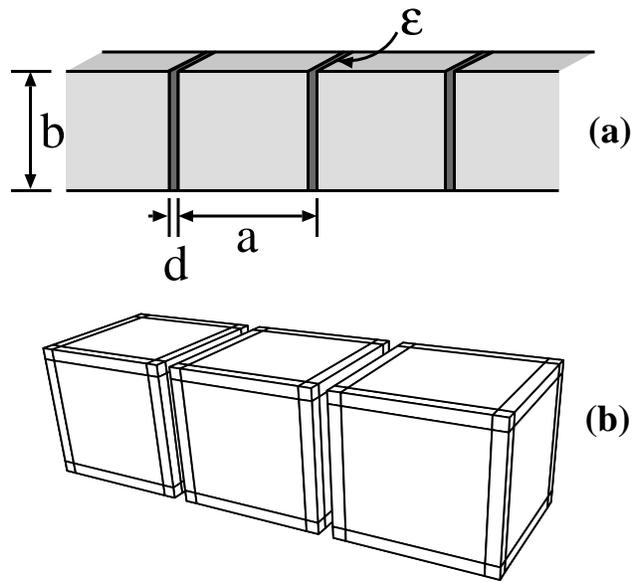
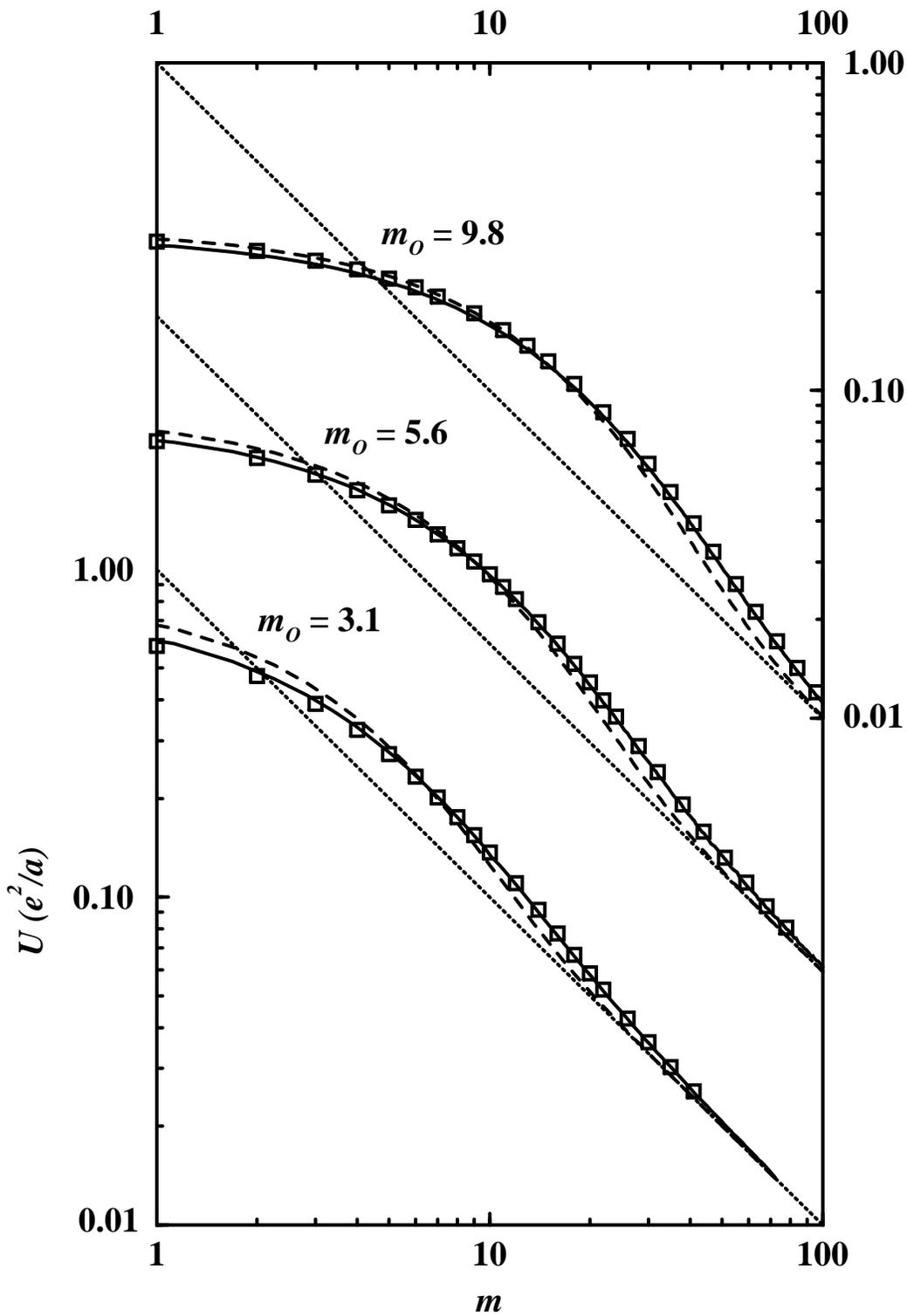
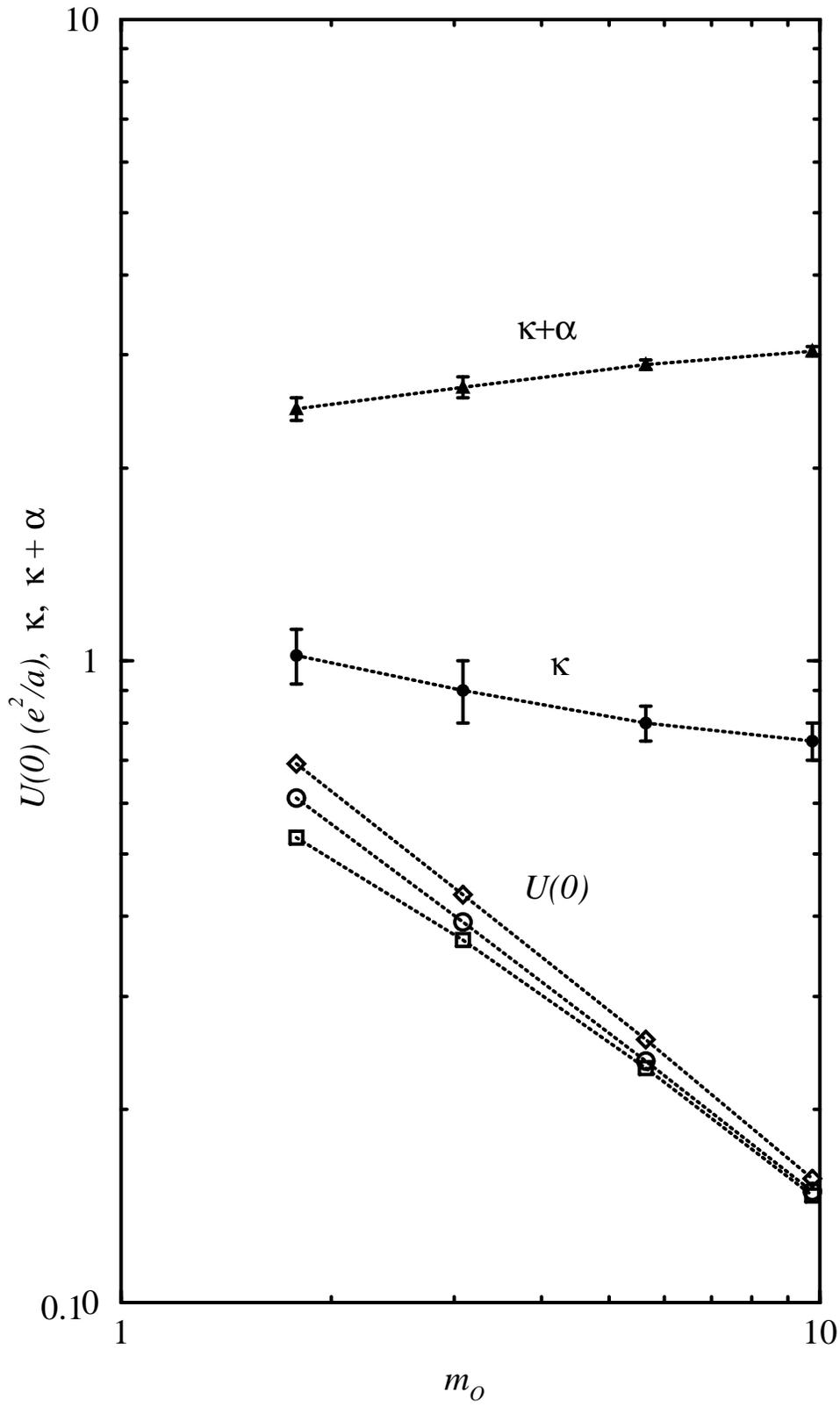
\end{document}